\documentclass[a4paper]{article}
\usepackage{todonotes}
\usepackage{INTERSPEECH2020}
\usepackage{multirow,url}
\usepackage{hyperref}

\usepackage{comment}
\newcommand{\dan}[1]{\texttt{#1}}

\title{Data Augmenting Contrastive Learning of Speech Representations\\ in the Time Domain}

\name{Eugene Kharitonov$^{*1}$, Morgane Rivière$^{*1}$, Gabriel Synnaeve$^1$, Lior Wolf$^1$, \\ Pierre-Emmanuel Mazaré$^1$, Matthijs Douze$^1$, Emmanuel Dupoux$^{1,2}$
  \thanks{* Contributed equally, order is chosen randomly.}
  }

\address{
  $^1$Facebook AI Research 
  $^2$EHESS, CNRS, INRIA, ENS-PSL University
}
\email{\{kharitonov,mriviere,gab,wolf,pem,matthijs,dpx\}@fb.com}

\begin{document}
\maketitle

\begin{abstract}
Contrastive Predictive Coding (CPC), based on predicting future segments of speech based on past segments is emerging as a powerful algorithm for representation learning of speech signal. However, it still under-performs other methods on unsupervised evaluation benchmarks. Here, we introduce WavAugment, a time-domain data augmentation library and find that applying augmentation in the past is generally more efficient and yields better performances than other methods. We find that a combination of pitch modification, additive noise and reverberation substantially increase the performance of CPC (relative improvement of 18-22\%), beating the reference Libri-light results with 600 times less data. Using an out-of-domain dataset, time-domain data augmentation can push CPC to be on par with the state of the art on the Zero Speech Benchmark 2017. We also show that time-domain data augmentation consistently improves downstream limited-supervision phoneme classification tasks by a factor of 12-15\% relative. 

\end{abstract}

\section{Introduction}

Recent works have demonstrated an interest in unsupervised representation learning as a pretraining method to obtain good speech features for downstream tasks with little labelled data \cite{schneider2019wav2vec,kahn2020,riviere2020multi,kawakami2020robust,wang2020}. While Contrastive Predictive Coding (CPC) and derivatives appear to be versatile methods for unsupervised representation learning \cite{oord2018representation,chung2019,chung2019generative}, they do not yet reach the state-of-the-art (SOTA) results on purely unsupervised learning metrics \cite{kahn2020,oord2018representation,versteegh2016,dunbar2017}. 

Data augmentation is useful for supervised training, and is also a key component in unsupervised setups in the image domain~\cite{dosovitskiy2014discriminative,chen2020simple}. It is not well established in unsupervised learning for speech, where the sequential nature of the signal may introduce specificities. 

Our first objective is to explore several types of time-domain data augmentation (additive noise, masking, reverberation) and several methods for augmenting in the contrastive framework (in the past, future, or both) in English (LibriSpeech). 
In a second stage, we extend the results to other languages (French and Mandarin) in the zero-resource 2017 benchmark~\cite{dunbar2017}. 
Lastly, we show that data augmentation benefits semi-supervised training, using the Libri-light benchmark~\cite{kahn2020}.

\section{Related work}

{\bf CPC.}
Van den Oord et al.~\cite{oord2018representation} introduced Contrastive Predictive Coding, a method for unsupervised representation learning. 
Applied to speech, CPC trains a convolutional encoder and a predictor for future embeddings of the encoder. 
To prevent mode collapsing, the loss is contrastive: an embedding should be close to positive future embeddings and distant from negative future embeddings.
CPC was used as pretraining for ASR~\cite{schneider2019wav2vec} and speaker identification~\cite{lowe2019,lai2019}.
Non-contrastive versions of predictive coding with fixed embeddings can learn generic multi-task representations~\cite{chung2019,chung2019generative}. 
Here we use a deeper and optimized version of the CPC implementation of ~\cite{kahn2020, riviere2020multi}.

\noindent
{\bf Data augmentation for ASR.}
Basic time-domain augmentations modify the sampling rate of the input by a small factor ($\pm 10\%$), which changes both the duration and pitch~\cite{ko2015audio}. 
Another one consists of adding noise, convolved with a room impulse response function to simulate point sources spread in space~\cite{ko2017rev}.
SpecAugment~\cite{park2019specaugment} is a spectral-domain augmentation whose effect is to mask bands of frequency and/or time. 
We introduce WavAugment, that implements these augmentations in the time domain and is optimized for applying augmentations on-the-fly as part of data loading.

Our work is close to~\cite{ravanelli2020multi}, which applies data augmentation techniques to representation learning (autoencoders). 
However, they evaluated them in terms of pretraining for a downstream task not in terms of the learned representation.


\vspace{-0.7em}
\section{Method}

Our method is based on a state-of-the-art CPC architecture~\cite{riviere2020multi}. We explore how to perform data augmentation and introduce the WavAugment package.

\vspace{-0.5em}
\subsection{The CPC2 architecture}
\label{ss:cpc_arch}

\begin{figure}[t]
  \centering
  \includegraphics[width=7.0cm]{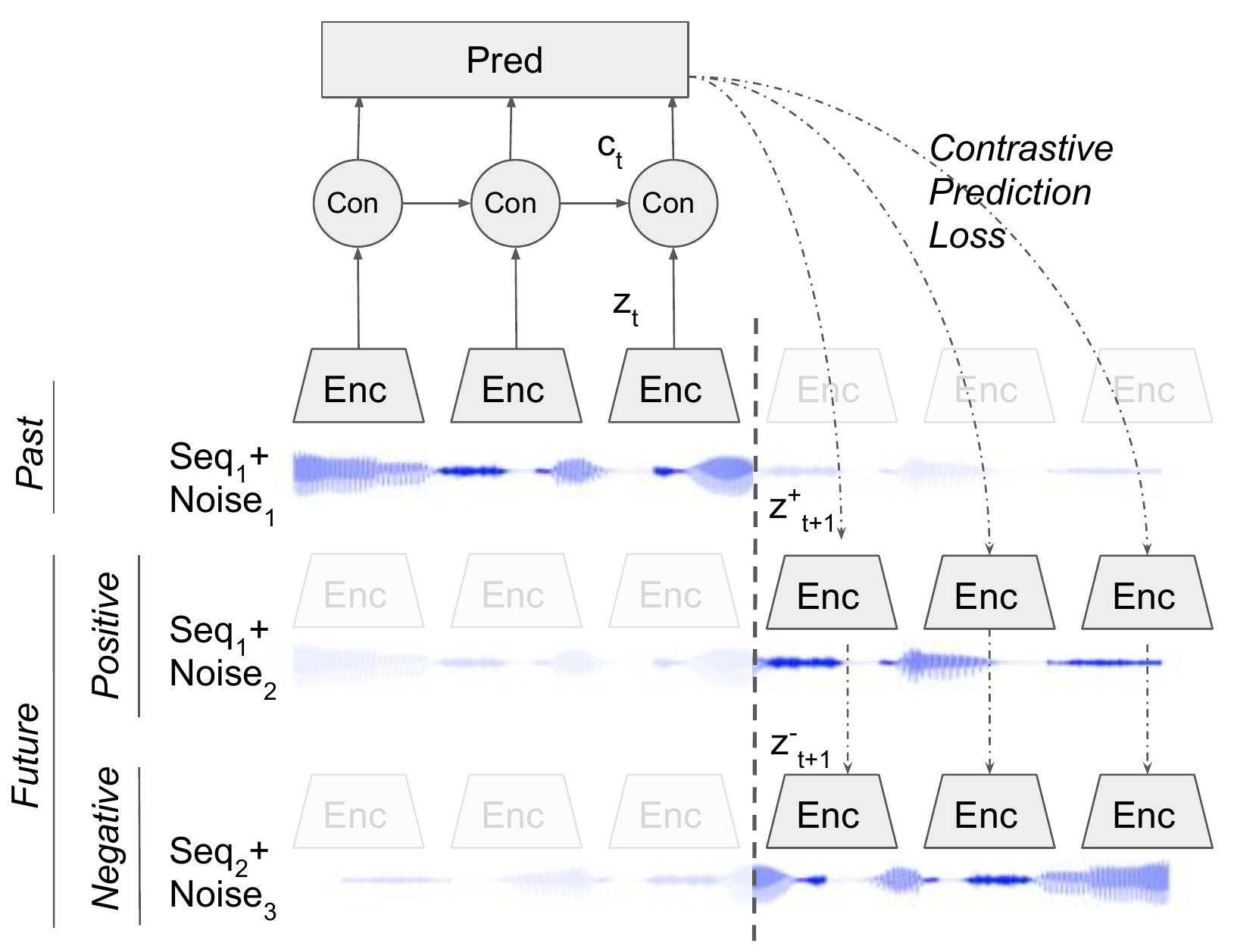}
  \vspace{-1.3mm}
\caption{\textbf{CPC-based data augmentation}. Each speech sequence is encoded twice, one for \textit{past} one for \textit{future}, with potentially different augmentations for each. The CPC loss tries to contrastively predict \textit{future} embeddings $z_{t+i}$ based on \textit{past} ones, ignoring the noise of the augmentation. Positive and negative sequences may have different augmentations.}
\label{fig:CPCaug}
\vspace{-1em}

\end{figure}

The architecture is summarized in Figure~\ref{fig:CPCaug}.
A convolutional encoder network produces a representation $z_t$ of the raw audio waveform. 
The sequence $(z_t)$ is then passed to a recurrent context network to build our final representation $c_t$. 
At each step, we apply $c_t$ to a predictor neural network $Pred$ with several outputs $Pred^k$ each one reconstructing future representations $z_{t+k}$ ($0< k \leq K$, $K=12$). The loss is contrastive and tries to minimize the dot product between the predicted and correct future representation while maximizing the dot product with a sample of 128 negative examples $\mathcal{N}_{t,k}$ taken from the batch. 
This gives the following loss:
\vspace{-0.3em}
\[\mathcal{L} = \frac{1}{K} \sum_{k=1}^K \log \frac{\exp ( Pred^k(c_t)^T z_{t+k})}{\sum_{n\in \mathcal{N}_{t,k}} \exp (Pred^k (c_t)^T z_{n})}\]
\vspace{-0.3em}

CPC2 is a modified version of the CPC architecture in~\cite{kahn2020,riviere2020multi}. The encoder
architecture is unchanged (5 convolutional layers with kernel sizes [10,8,4,4,4], strides [5,4,2,2,2] and hidden dimension 256). We increase the depth of the auto-regressive network, which improves accuracy (see Supplementary Table \ref{tab:archi_ablation})  
For the recurrent context nextwork, we use a 2-layer LSTM, as a tradeoff between feature quality and training speed.
In the prediction network, we replace the $k$ independent transformers in \cite{kahn2020,riviere2020multi}, each one predicting a specific time-step ahead, to a single multi-head transformer layer with $k$ classifiers at its heads. 
This has a limited impact on accuracy but dramatically decreases training time.


\subsection{Data augmentation and CPC}
\label{ss:past_future}

As discussed in Section~\ref{ss:cpc_arch}, the encoded representations $z_t$ are used in two ways: 
(a) to calculate the contextual representation $c_t$
, and 
(b) as target predictions (positive or negative candidates).
 We refer to the representation $z_t$ as \textit{past} and the targets $z^+, z^-$ as \textit{future}.
The model predicts \textit{future} representations  based on its \textit{past}, by learning to discriminate it from a samples of negative candidates (Figure~\ref{fig:CPCaug}).

We can apply two different augmentations on the same speech sequence and use them to calculate \textit{past} and \textit{future} representations. 
This separation opens a plethora of possibilities for data augmentation:
applying the same augmentation on all sequences in the batch (on query sequence and all positive and negative candidates); 
augmenting each sequence independently (\textit{past} and \textit{future} have identical augmentations, but negatives have independent augmentations); 
augmenting only \textit{past}; 
augmenting only \textit{future}; 
augmenting both \textit{past} and \textit{future} independently (\textit{past+future} setting).
Preliminary experiments demonstrated that the most promising approaches are either augmenting only the \textit{past} representation or applying two independent augmentations on both \textit{past} and \textit{future} (\textit{past+future}).
In this work, we therefore focus on these two options.

\subsection{WavAugment}

Our experimental setup requires to apply independent data augmentations on short  audio sequences ($\approx 1$ s). For this, we developed WavAugment, a library that implements time-domain augmentations. WavAugment is publicly available at~{\color{blue}\href{https://github.com/facebookresearch/WavAugment}{github.com/facebookresearch/WavAugment}}. 
WavAugment builds upon a C++ API to \textit{libsox}\footnote{http://sox.sourceforge.net/sox.html} that implements dozens of audio processing transformations. WavAugment has a Pytorch~\cite{Pytorch2019} interface and Pytorch- and libsox-based effects can be interleaved transparently.

\subsection{Datasets and evaluation measures}\label{sec:dataset}


In the experiments reported below, unless reported otherwise, we trained our CPC model on Librispeech-100h~\cite{librispeech}, which is a set of short sentences in good quality (clean) read speech, from a balanced set of speakers. We directly used all of the files, without filtering or modification. In Experiment~2, we introduce two similar datasets in French and Mandarin, respectively. The French dataset was created by selecting the French data from the Librivox website\footnote{\url{https://librivox.org/}}, and Mandarin from the MagicData dataset~\cite{magicdata}. The recordings were cut into ``utterance''-like segments using pyannote's  Voice Activity Detector \cite{Bredin2020}. Both datasets had a similar number of speakers and total duration as Libri-Speech (250 speakers, 76h and 80h respectively). 

We tested the learned representation using the Libri-light~\cite{kahn2020} ABX metric for unsupervised representation learning. This distance-based metric estimates the probability that a speech segment $X$ is closer to a segment $A$ with the same transcription than to a segment $B$ with a different transcription. The distance is the DTW-realigned average angle (arc-cosine of the normalized dot product) between each frames
. The test uses minimal pairs of triphones that only change in the central phoneme ('bet' vs 'bit'), and is conducted within-speaker ($A$, $B$ and $X$ are from the same speaker) and across-speaker ($A$ and $B$ are from one speaker, $X$ from another one). This metric has been shown to be useful to analyse the linguistic content of speech features without having to train a classifier \cite{schatz2016abx}, and has been used in the Zero Resource Challenge series \cite{versteegh2016,dunbar2017,dunbar2019}.   
\section{Experiments}

\subsection{Preliminary Experiment: Tuning data augmentation}

We focus on 5 augmentations that were either proposed earlier~\cite{park2019specaugment} or that can potentially inject useful invariances in the speech representations. 
We selected: pitch modification (\dan{pitch}), additive noise (\dan{add}), reverberation (\dan{reverb}), band reject filtering (\dan{bandrej}), and time masking (\dan{tdrop}). The last two augmentations are similar to those used in SpecAugment~\cite{park2019specaugment}.
The \dan{pitch} can be attributed to the \textit{source} (how the speaker talks), \dan{add} and \dan{reverb} to the \textit{communication channel}, and \dan{bandrej} \& \dan{tdrop} to noise in the \textit{neural representation} of the speech. 
When we compose augmentations (indicated by '+'), they are applied in that order.

\label{ss:design}
In pilot experiments we calibrated the strength of the augmentations looking at the overall ABX results (within and across on the dev clean and other set of Libri-light \cite{kahn2020}. 
For \dan{pitch}, the applied change in the pitch is an integer sampled uniformly between +300 and -300 (the change value is measured by 1/100 of a tone). 
In \dan{reverb}, we uniformly sample \textit{room-scale} between 0 and 100, fixing other parameters to defaults. 
\dan{tdrop} zeroes out one random subsequence of length 50ms. 
We found that \dan{bandrej} performs best when we set the maximal width of the rejected spectrum to $150$ Hz. 

We discovered accidentally 
that for additive noise, low frequencies are more effective than high frequencies. 
We therefore explored systematically the effect of the spectral characteristics of noise by filtering sounds from the MUSAN dataset~\cite{musan2015} 
in successive frequency bands. We selected 5 broad bands, defined by 4 cutoff points by the tripling of the frequency: 80Hz, 240Hz, 720Hz, 2160Hz). 
We found that the optimal additive noise was obtained by bandpass filtering MUSAN sounds in the $[80,240]$ Hz range, which corresponds roughly 
the human F0 (see Supplementary Table~\ref{tab:add}).
\begin{table}[]
\begin{center}
\begin{tabular}{l c @{\hspace{0.8\tabcolsep}} c @{\hspace{0.8\tabcolsep}} c@{\hspace{0.5\tabcolsep}}c @{\hspace{0.8\tabcolsep}} c @{\hspace{0.8\tabcolsep}}  }
\toprule
  & \multicolumn{2}{c}{{Within spk.}}  && \multicolumn{2}{c}{{Across spk.}} \\
\cline{2-3}\cline{5-6}
                          &dev       &dev   && dev  &dev\\
System                    &clean     &other && clean&other\\
\midrule
MFCC Baseline           &   10.95 &  13.55 &&  20.94 & 29.41 \\
CPC LL-60k  \cite{kahn2020} &    6.11 &   8.17 &&   8.05 & 12.83 \\ 
\midrule
\midrule
\multicolumn{6}{l}{\emph{Single augmentations (CPC2 on LibriSpeech clean 100h)}}\\
no augmentation     &    6.06 &   8.18 &&    7.59 &    12.84 \\ 
\midrule
pitch-past          &    4.90 &   6.28 &&	 6.84 &	   11.04 \\ 
pitch-past+future   &    5.03 &	  6.35 &&	 7.11 &	   11.30 \\ 
\midrule
add-past            &    5.47 &   7.58 &&    6.97 &    12.17 \\ 
add-past+future     &    5.16 &   7.33 &&    6.77 &    11.71 \\ 
\midrule
reverb-past         &    5.55 &   7.61 &&    7.16 &    12.19 \\ 
reverb-past+future  &    5.58 &   7.91 &&    7.77 &    13.07 \\ 
\midrule
bandrej-past        &    5.83 &   7.88 &&    7.07 &    12.21 \\ 
bandrej-past+future &    5.92 &   7.81 &&    7.19 &    12.24 \\ 
\midrule
tdrop-past         &    5.78 &   7.92 &&    7.18 &    12.56 \\ 
\midrule
\midrule
\multicolumn{6}{l}{\emph{2-way combinations, past only (same model and train set)}}\\
pitch+add          &  4.81 & 6.03 && 6.79 & 10.90 \\
pitch+reverb       &  4.74 & 6.75 && \textbf{6.06} & 10.99 \\ 
pitch+tdrop        &  4.83 & 6.15 && 6.90 & 11.08  \\ 
add+reverb         &  5.41 & 6.87 && 7.41 & 11.97 \\
add+tdrop          &  5.38 & 6.97 && 7.70 & 12.22 \\
reverb+tdrop       &  5.41 & 6.93 && 7.32 &	12.05  \\ 
\midrule
\multicolumn{6}{l}{\emph{3-way combinations, past only (same model and train set)}}\\
pitch + add + reverb    &\bf 4.66 &\bf 5.81 &&   6.62 &\bf 10.60 \\
pitch + add  + tdrop    &    4.86 &    6.09 &&   6.70 &    10.78 \\
pitch + reverb + tdrop  &    4.72 &    6.02 &&   6.53 &    10.70 \\
add + reverb + tdrop    &    5.40 &    6.87 &&   7.47 &    11.98 \\
\midrule
\multicolumn{6}{l}{\emph{4-way Combinations, past only (same model and train set)}}\\
pitch+add+reverb+tdrop  &    4.87 &    6.08 &&   6.79 &    10.76 \\ 
\bottomrule
\end{tabular}
\caption{\textbf{ABX errors on data-augmented CPC features (Libri-light dev set).} Within- and across-speaker phoneme discriminability scores (lower is better) on the Libri-light clean and other dev sets for CPC training as a function of types of data augmentation, in isolation or combination (see Section~\ref{ss:design}).
\label{tab:effects}
}
\end{center}\vspace{-20pt}
\end{table}

\vspace{-5pt}
\subsection{Experiment 1: Data augmentation combinations}
\label{sec:daeffects}
\vspace{-1pt}

We first tested these five augmentations alone, either applying them to the \textit{past} of the sequence or independently to \textit{past} and \textit{future} (\textit{past+future}) (see Section~\ref{ss:past_future}). 

On analyzing single augmentations in Table~\ref{tab:effects}, we first observed that in many cases applying augmentations on \textit{past} performs as well as, or even better, than \textit{past+future} (\dan{pitch}, \dan{reverb}, \dan{bandrej}). The only augmentation performing better on \textit{past+future} is \dan{add}.\footnote{We did not experiment with \dan{tdrop} applied on \textit{past+future} as this will zero out the predicted sequences.}  
According to their average performance, the individual augmentations can be sorted, from most to least useful: \dan{pitch}, \dan{add}, \dan{reverb}, \dan{tdrop}, and  \dan{bandrej}.

Next, we study the performance of combinations of augmentations. We decided to drop \texttt{bandrej} from consideration due to its poor results. We only consider augmenting \textit{past}, as this gives roughly the same quality of representations, but requires less computation. 
As a result, we have 6 possible two-way, 4 three-way, and 1 four-way combination of effects. 
The results are in the lower part of Table~\ref{tab:effects}, and they show that \texttt{pitch+add+reverb} performs best in 3 out of 4 metrics.

We chose this combination and evaluated the corresponding model on the Libri-light test set. The results are reported in Table~\ref{tab:best} and show that, across all metrics, data augmentation yields relative improvements of 18-22\% over no augmentation, and ends up with better results than the original CPC algorithm run on the much larger 60k hours dataset. 

\begin{table}[]
\begin{center}
\begin{tabular}{l c @{\hspace{0.8\tabcolsep}} c @{\hspace{0.8\tabcolsep}} c@{\hspace{0.5\tabcolsep}}c @{\hspace{0.8\tabcolsep}} c @{\hspace{0.8\tabcolsep}}  }
\toprule
  & \multicolumn{2}{c}{{Within spk.}}  && \multicolumn{2}{c}{{Across spk.}} \\
\cline{2-3}\cline{5-6}
                          &test       &test  && test&test\\
System                    &clean     &other && clean&other\\
\midrule
MFCC Baseline       &  10.58    & 13.60    && 20.45 &  28.5  \\
CPC LL-60k   \cite{kahn2020}& 5.83  & 8.14  && 7.56&  13.42 \\ 
\midrule
\multicolumn{4}{l}{\emph{Trained on Librispeech-100h}}\\
CPC2        &  5.69     &   8.42   &&   7.26   &   13.42 \\ 
CPC2+WavAug        &    \bf 4.46   &  \bf 6.56   && \bf 5.90 & \bf 10.95 \\
\bottomrule
\end{tabular}
\caption{\textbf{ABX errors on data-augmented CPC features (Libri-light test sets).} Within- and across-speaker phoneme discriminability scores (lower is better) on the Libri-light test sets for our best augmentation (\texttt{pitch}+\texttt{add}+\texttt{reverb}-\texttt{past}).}\label{tab:best}
\end{center}\vspace{-20pt}
\end{table}

\vspace{-5pt}
\subsection{Experiment 2: Extending to other languages}
\vspace{-1pt}

In this experiment, we tested whether our data augmentation technique could be extended to other languages. We selected the three dev datasets of the ZeroSpeech Challenge 2017, covering English, French, and Mandarin. 
As in the previous experiment, the metrics are the within- and across- ABX test provided by the Challenge. 
For training, we used both the small in-domain training sets provided by the Challenge (45h, 24h, and 2h30, respectively), and our own, larger, out-of-domain training sets. For English, we used Librispeech-100 (100h), for French, the 76h of French-librivox, and Mandarin, the 80h of MagicData described in Section \ref{sec:dataset}. We also observed, training on Librispeech-100 and testing on Libri-light dev, that using larger datasets in combination with data augmentation allowed to benefit from increasing the number of LSTM layers to 3 (see Supplementary). We included this modification in the experiments.

The results are shown in Table \ref{tab:zr}. As can be seen, while noise augmentation improves the score on all three languages, we cannot reach the SOTA with the small training datasets provided from the challenge. We can however, be on par with or improve  over best performing baseline with our out-of-domain train sets (same languages, larger datasets), in particular with the larger model. This shows that while our technique scales with dataset size, it is still less data efficient than the techniques described in Heck et al.~\cite{heck2017} and Choroskwi~et~al.~\cite{chor2019}. Note however, that both studies used speaker adaptation which are outside the scope of what can be done with standard time domain data augmentation techniques. 

\begin{table}[h]
\centering
\begin{tabular}{l@{\hspace{0.7\tabcolsep}} c@{\hspace{0.7\tabcolsep}}c@{\hspace{0.7\tabcolsep}} c@{\hspace{0.3\tabcolsep}} c@{\hspace{0.7\tabcolsep}}c@{\hspace{0.7\tabcolsep}} c@{\hspace{0.3\tabcolsep}} c@{\hspace{0.7\tabcolsep}}c@{\hspace{0.7\tabcolsep}}c}
\toprule
& \multicolumn{2}{c}{English} && \multicolumn{2}{c}{French} && \multicolumn{2}{c}{Mandarin} \\
\cmidrule{2-3}\cmidrule{5-6}\cmidrule{8-9}
                                    & W. & A.     && W.    & A.    && W. & A.  &AVG\\
\midrule
\multicolumn{9}{l}{\emph{Trained on ZeroSpeech2017 (45h, 24h, 2h30, resp.)}}\\
Superv. topline~\cite{dunbar2017}   &   5.3 &    6.9 &&   6.8 &   9.1 &&    4.2 &   5.7 &6.33\\
Heck et al.~\cite{heck2017}         &   6.2 &    8.7 &&   8.7 &  11.7 &&  7.9 &\bf 7.4&8.43\\
Chorow. et al.~\cite{chor2019}    &   5.5 &    8.0 &&\bf 7.5&\bf 10.8&&   10.7 &  11.2 &8.95\\ 
\midrule
CPC2                                &   8.6 &   12.0 &&  12.2 &   16.4 &&  12.0 &  14.0 &12.53 \\ 
CPC2+WavAug                         &   6.6 &    9.3 &&   9.3 &   14.1 &&  11.2 &  11.9 &10.4\\ 
\midrule
\midrule
\multicolumn{9}{l}{\emph{Trained on out-of-domain (100h, 76h, 80h, resp.)}}\\
CPC2                                &   6.1 &    8.7 &&  10.3 &   12.9 &&   9.3 &   9.6 &9.48\\ 
CPC2+WavAug                         & 4.7& 6.5 &&   8.6 &   11.1 &&  7.9&   7.8 &  7.77\\ 
CPC2-3L+WavAug                   &\bf 4.6 &\bf 5.8 &&   7.6 &   10.9 && \bf 7.8 &   8.0 &\bf 7.45 \\ 
\bottomrule
\end{tabular}
\caption{
\textbf{ABX errors on the ZeroResource Speech Challenge 2017 (120s).}
Within- (``W.'') and across-speaker (``A.'') phoneme discriminability scores on English, French and Mandarin speech for CPC features with and without data augmentation. For comparison, the best systems plus supervised topline of the ZeroSpeech leaderboard trained on the provided datasets.}
\label{tab:zr}
\vspace{-10pt}
\end{table}

\vspace{-5pt}
\subsection{Experiment 3: Pretraining and limited supervision}
\vspace{-1pt}

In this experiment, we test whether our data augmentation technique can build better speech features that can be used for downstream tasks. 
Here, we use the Libri-light limited supervision phone classification task \cite{kahn2020}, which contains intentionally small training sets (10 min, 1h or 10 hours of labelled data). 
We fine-tune a linear phone classifier built on top of the CPC features with a CTC loss (frozen features). On 10 hours of data, we also fine-tune the entire network. Again, we additionally experiment with an architecture that has a 3-layer LSTM (CPC2-L3) (See Supplementary Table \ref{tab:archi_search}).

The results are in Table \ref{tab:semi} and show an effect of signal-based data augmentation, both for pretraining and for fine tuning. 
For the supervised fine-tuning phase, we found out that we got the best results by using only pitch augmentation. 
 Other methods having low or negative effects in this case. The combined effects of data augmentation on pretraining and fine-tuning adds up to 12-15\% relative improvement across the different training sets.  
Interestingly, we find that with data augmentation we can beat the reference baseline (pretraining on 60k hours plus fine tuning on 10 hours) on frozen features with substantially less data (pretraining on 100 hours, plus fine tuning on 1 hour). 
Another point worth mentioning is that with data augmentation, 10 minutes of data on frozen features is sufficient to outperform the no-pretraining reference 
with 10 hours of labels.

\begin{table}[h]
\begin{tabular}{l cc @{\hspace{0.8\tabcolsep}} c @{\hspace{0.8\tabcolsep}} c @{\hspace{0.8\tabcolsep}} c }
\toprule
                     &Augmented& dev-      & dev-    & test-     & test-\\
System               &fine-tuning& clean     & other   & clean     & other\\
\midrule
\multicolumn{5}{l}{\emph{Reference}}\\
\multicolumn{2}{l}{CPC unlab-60k+train-10h-full} & 28.4& 41.4 & 27.9 & 43.6 \\
\multicolumn{2}{l}{CPC no pretraining - 10h-full}  & 45.9& 55.7 & 43.7 & 58.6 \\
\multicolumn{2}{l}{CPC2 no pretraining - 10h-full} & 41.3& 52.3 & 39.3 & 56.1 \\
\midrule
\midrule
\multicolumn{5}{l}{\emph{Frozen features - classifier trained on 10min}}\\
\multirow{2}{*}{CPC2}       &No & 47.8 & 60.9 & 47.0 & 60.1\\
                            &Yes& 49.4 & 57.9 & 49.4 & 59.2\\
\multirow{2}{*}{CPC2+WavAug}&No & 39.5 & 51.3 & 39.1 & 52.4\\
                            &Yes& 41.6 & 51.7 & 41.7 & 52.9\\
\midrule
\multicolumn{5}{l}{\emph{Frozen features - classifier trained on 1h}}\\
\multirow{2}{*}{CPC2}       &No & 34.6 & 47.5 & 32.9 & 50.0\\
                            &Yes& 33.5 & 46.9 & 32.7 & 49.4\\
\multirow{2}{*}{CPC2+WavAug}&No & 29.1 & 42.4 & 28.8 & 44.3\\
                            &Yes& 28.0 & 41.3 & 27.8 & 43.3\\
\midrule
\multicolumn{5}{l}{\emph{Frozen features - classifier trained on 10h}}\\
\multirow{2}{*}{CPC2}       &No & 29.3 & 43.7& 29.0 & 47.1\\
                            &Yes& 31.1 & 44.9& 30.6 & 48.3\\
\multirow{2}{*}{CPC2+WavAug}&No & 26.1 & 39.9& 25.7 & 41.6\\
                            &Yes& 25.7 & 39.3& 25.3 & 41.2\\
\midrule
\multicolumn{5}{l}{\emph{Full fine-tuning, 10h of data}}\\
\multirow{2}{*}{CPC2}       &No & 27.8 & 42.6 & 26.5 & 45.0\\
                            &Yes& 26.3 & 39.9 & 25.4 & 43.9\\
\multirow{2}{*}{CPC2+WavAug}&No & 24.5 & 39.0 & 24.1 & 40.8\\
                            &Yes& 23.5& 37.6& 23.1&41.0\\
\multirow{2}{*}{CPC2-L3+WavAug}&No & 22.9 & 37.3 & 22.8 & \bf{39.9}\\
                            &Yes&\bf 22.5 &\bf 36.8 &\bf 22.2& \bf{39.9}\\
\bottomrule
\end{tabular}
\caption{\textbf{Phone Error Rate (PER) in the semi-supervised setting.} A linear classifier is added on top of Librispeech-100 pretrained CPC2 models and fine tuned with either 10min, 1h or 10h of Libri-light labelled data with a CTC loss.  For comparison, reference Libri-light results plus the untrained CPC2 architecture fully fined-tuned with 10 h. 
} 
\label{tab:semi}
\vspace{-10pt}
\end{table}

\section{Discussion}
We have introduced WavAugment, a library for time-domain data augmentation and illustrated its use in the context of unsupervised contrastive representation learning, and in the context of learning with limited supervision. We found that pitch and additive noise are the most powerful data augmentation techniques for our implementation of contrastive predictive coding, yielding very good results in unsupervised representation learning in English, Mandarin and French. We further showed that these gains extend to fine tuning on very limited data yielding gains in PER. Interestingly, the two most popular data augmentation techniques that are typically done in the spectral domain (as in SpecAugment)  do not work very well for CPC training. Furthermore, pitch and additive noise are techniques that can only be applied in the time domain. Further work will allow to determine whether the superiority of time domain noise augmentation over spectral ones is specific to the CPC loss or to the fact that our architecture starts directly from the waveform as opposed to using spectral features like Mel Filterbanks or MFCCs. Note that~\cite{ravanelli2020multi} also combines several data augmentation techniques for unsupervised learning in an autoencoder architecture. Among data augmentation technique they use the most are two time-domain ones (reverberation and additive noise) and one spectral (band reject). It remains to be seen how pitch would fare in such a pretraining setup. 

\section{Conclusion}
With data augmentation, CPC can take good advantage of relatively short (around 100 hours) clean and well segmented speech, although it is currently insufficient to learn competitively with very small amounts of data (between 2.5 and 50 hours). More research is needed to extend such techniques in both directions: with small amounts of data, and with very large, and potentially more noisy datasets.  In addition, the differences that we observe between data-augmentation effects open the issue of more systematic exploration of data augmentation as a function of tasks and architectures. 



\let\oldbibliography\thebibliography
\renewcommand{\thebibliography}[1]{%
  \oldbibliography{#1}%
  \setlength{\itemsep}{0pt}%
}
\vspace{-5pt}

\bibliographystyle{IEEEtran}
\bibliography{mybib}


\clearpage
\setcounter{section}{0}
\setcounter{table}{0}
\setcounter{figure}{0}

\renewcommand\thesection{S\arabic{section}}
\renewcommand\thetable{S\arabic{table}}
\renewcommand\thefigure{S\arabic{figure}}

\section{Supplementary Results}\label{sec:supmeth}
\subsection{Changing the architecture: ablation study}

We started from the model described in \cite{riviere2020multi}: the encoder network is composed of 5 convolutional layers with kernel sizes [10,8,4,4,4], strides [5,4,2,2,2] and hidden dimension 256. 
We worked with ReLU activation and inserted a channel normalization procedure between each convolutional layer.
As far as the context network is concerned, we used a 2-layers LSTM. Finally, we used a single layer multihead transformer to do the prediction instead of several single head transformers. Table \ref{tab:archi_ablation} shows different ablations that we ran to compare these different versions. 

We ran our experiments using the Adam optimizer with $lr=2e-4, \beta_1 = 0.9, \beta_2 = 0.999$. 
Although we didn't resort to learning rate decay, we used a learning rate ramp for the first $10$ epochs.

\begin{table}[h]
\begin{center}
\begin{tabular}{l c @{\hspace{0.8\tabcolsep}} c @{\hspace{0.8\tabcolsep}} c@{\hspace{0.5\tabcolsep}}c @{\hspace{0.8\tabcolsep}} c @{\hspace{0.8\tabcolsep}}  }
\toprule
  & \multicolumn{2}{c}{{Within spk.}}  && \multicolumn{2}{c}{{Across spk.}} \\
\cline{2-3}\cline{5-6}
                      &dev       &dev   && dev  &dev\\
System                &clean     &other && clean&other\\
\midrule
CPC LS-100 \cite{riviere2020multi} & 6.81   &  8.91   &&  8.46  & 13.70    \\
\midrule
CPC + 2 layers LSTM   &    5.97 &    8.12  &&    7.39 &    12.79  \\ 
CPC + 3 layers LSTM   &    5.93 &  8.41   &&    7.76 &    13.14  \\ 
\midrule
CPC + MH              &    6.84 &    9.10  &&    8.68 &    14.08  \\ 
\midrule
CPC + MH +  2 layers LSTM & \textbf{6.02} &    \textbf{8.11}  &&    \textbf{7.56} &   \textbf{12.91}  \\
\bottomrule
\end{tabular}
\caption{\textbf{Architecture ablations, ABX errors (Libri-light dev set)}. We compare the original CPC model described in \cite{riviere2020multi} with modifications including more LSTM layers and a single Multi-Head prediction model for all time steps (MH). The bottom model is the one we refer to as CPC2 in the paper.}\label{tab:archi_ablation}
\end{center}\vspace{-20pt}
\end{table}

\begin{table}[]
\begin{center}
\begin{tabular}{l c @{\hspace{0.8\tabcolsep}} c @{\hspace{0.8\tabcolsep}} c@{\hspace{0.5\tabcolsep}}c @{\hspace{0.8\tabcolsep}} c @{\hspace{0.8\tabcolsep}}  }
\toprule
  & \multicolumn{2}{c}{{Within spk.}}  && \multicolumn{2}{c}{{Across spk.}} \\
\cline{2-3}\cline{5-6}
                      &dev       &dev   && dev  &dev\\
System                &clean     &other && clean&other\\
\midrule
\textit{3h Libri-light}  \\
CPC2 +  2 layers LSTM & 12.82 & 13.81 &&	17.21  & 20.85 \\ 
CPC2 + 3 layers LSTM  & 13.22 & 14.30 &&  18.30 & 22.27 \\
\midrule
\textit{45h Libri-light}  \\
CPC2 + 2 layers LSTM & 6.31	 & 8.37 && 8.52 & 13.38 \\ 
CPC2 +  3 layers LSTM  & 6.38	& 8.29 && 8.43 & 13.72 \\ 
\midrule
\textit{100h LibriSpeech}  \\
CPC2 + 2 layers LSTM &  4.66 & 6.62  && 5.81 & 10.60 \\ 
CPC2 +  3 layers LSTM  & 4.24 & 6.38 &&  5.76 & 10.43 \\ 
\midrule
\bottomrule
\end{tabular}
\caption{\textbf{Architecture ablations, ABX errors (Libri-light dev set)}. We compare modifications of the CPC architecture across different dataset sizes. In all cases, we apply the best data augmentation reported in the main text.}\label{tab:archi_search}
\end{center}\vspace{-20pt}
\end{table}

\subsection{Changing the architecture: dataset size in presence of data augmentation}
In the next experiment, we study the performance of our model in function of the size of the available data  and the architecture size (controlled by the number of LSTM layers). We simulate the amounts of data available at ZeroSpeech2017 for Mandarin (3h), French (45h), and English (100h) by sub-sampling from LibriLight (3h and 45h) and using LibriSpeech (100h). In all experiments, we use the best data augmentation found in the main text (\texttt{pitch+add+reverb-past}). We report the obtained results in Table~\ref{tab:archi_search}.

We observe that in the cases of 3h and 45h datasets, the architecture with 2 layers of LSTM still perform best. However, with 100h of data, increasing the model depth turns out to be beneficial. On comparing with the results reported in Table~\ref{tab:archi_ablation}, we see that it is the presence of the data augmentation that allows us to leverage a deeper architecture.

\begin{table}[]
\begin{center}
\begin{tabular}{l c @{\hspace{0.8\tabcolsep}} c @{\hspace{0.8\tabcolsep}} c@{\hspace{0.5\tabcolsep}}c @{\hspace{0.8\tabcolsep}} c @{\hspace{0.8\tabcolsep}}  }
\toprule
  & \multicolumn{2}{c}{{Within spk.}}  && \multicolumn{2}{c}{{Across spk.}} \\
\cline{2-3}\cline{5-6}
                      &dev       &dev   && dev  &dev\\
System                &clean     &other && clean&other\\
\midrule
MFCC Baseline         & 10.95   &  13.55   &&  20.94  & 29.41    \\
CPC LL-60k            &   6.11 &  8.17   &&  8.05 &  12.83\\ 
\midrule
\midrule
\multicolumn{5}{l}{\emph{CPC2 -- Trained on LibriSpeech clean 80h}}\\
no augmentation       &    6.06 &    8.18  &&    7.59 &    12.8  \\ 
\midrule
\multicolumn{5}{l}{\emph{Band pass -- Musan -- past only}}\\
no filtering          &    5.81 &    7.40  &&    8.03 &    12.7  \\ 
$[0,80]$ Hz           &    5.55 &    7.56  &&    6.82 &    12.0  \\ 
$[80,240]$ Hz         &    5.38 &    7.58  &&    6.99 &    12.1  \\ 
$[240,720]$ Hz        &    6.22 &    8.32  &&    7.89 &    12.9  \\ 
$[720,2160]$ Hz       &    6.71 &    9.11  &&    8.52 &    13.8  \\ 
$[2160,8000]$ Hz      &    6.64 &    8.74  &&    8.30 &    13.4  \\ 
\midrule
\multicolumn{5}{l}{\emph{Band pass -- Musan -- past + future}}\\
no filtering          &    6.52 &    8.79   &&   8.20 &    13.5  \\ 
$[0,80]$ Hz           &    5.28 &    7.48  &&    6.83 &    12.1  \\ 
$[80,240]$ Hz         &    \bf 5.16 & \bf 7.33  &&  \bf 6.77 & \bf 11.7  \\ 
$[240,720]$ Hz        &    6.01 &    8.36  &&    7.45 &    12.9  \\ 
$[720,2160]$ Hz       &    7.40 &    9.83  &&    9.06 &    14.2  \\ 
$[2160,8000]$ Hz      &    7.40 &    9.72  &&    9.00 &    14.2  \\ 
\bottomrule
\end{tabular}
\caption{\textbf{Additive noise augmented CPC, ABX errors (Libri-light dev set).} Within- and across-speaker phoneme discriminability scores (lower is better) on the Libri-light clean and other dev sets for CPC training as a function of varying types of additive noise augmentation.}\label{tab:add}
\end{center}\vspace{-10pt}
\end{table}

\subsection{Frequency sensitive additive noise}
Here, we explore how frequency filtering affects additive noise data augmentation. We did two experiments: band-pass filtering, and lowpass filtering. For bandpass, here are the frequency bands we applied to the MUSAN dataset: $[0,80]$ Hz, $[80,240]$ Hz, $[240,720]$ Hz, $[720,2060]$ Hz, $[2160-8000]$ Hz. The second band corresponds roughly to the range of human pitch (F0), the third, to the range of the first formant (F1), the fourth to the range of the second formant (F2). The extreme ranges (very low or very high frequencies) do not typically carry much information. Table  \ref{tab:add} shows the effect of filtering in these bands before adding the noise to the speech signal. An optimal range seems to be $[80,240]$ Hz. For lowpass, we selected sucessive 100Hz bands, starting from zero. 

\end{document}